\documentclass[aps,prl,twocolumn,superscriptaddress,longbibliography]{revtex4-1}

\usepackage{amsmath}
\usepackage{epsfig}
\usepackage{graphicx}
\usepackage{graphicx, color, epstopdf}
\usepackage{bm}
\usepackage{amssymb}
\usepackage{hyperref}
\usepackage{xcolor}
\usepackage{subfigure}
\begin{document}
\bibliographystyle{apsrev4-1}
\title{A Non-Hermitian Moir\'{e} Valley Filter}
\author{Kai Shao}
\affiliation{National Laboratory of Solid State Microstructures, School of Physics, and Collaborative Innovation Center of Advanced Microstructures, Nanjing University, Nanjing, 210093, China}

\author{Hao Geng}
\affiliation{National Laboratory of Solid State Microstructures, School of Physics, and Collaborative Innovation Center of Advanced Microstructures, Nanjing University, Nanjing, 210093, China}

\author{Erfu Liu}
\affiliation{National Laboratory of Solid State Microstructures, School of Physics, and Collaborative Innovation Center of Advanced Microstructures, Nanjing University, Nanjing, 210093, China}

\author{Jose L. Lado}
\affiliation{Department of Applied Physics, Aalto University, 02150 Espoo, Finland}

\author{Wei Chen}
\email{Corresponding author: pchenweis@gmail.com}
\affiliation{National Laboratory of Solid State Microstructures, School of Physics, and Collaborative Innovation Center of Advanced Microstructures, Nanjing University, Nanjing, 210093, China}

\author{D. Y. Xing}
\affiliation{National Laboratory of Solid State Microstructures, School of Physics, and Collaborative Innovation Center of Advanced Microstructures, Nanjing University, Nanjing, 210093, China}
\begin{abstract}
A valley filter capable of generating a valley-polarized current is a crucial element in
valleytronics, yet its implementation remains challenging.
Here, we propose a valley filter made of a graphene bilayer
which exhibits a 1D moir\'{e} pattern in the overlapping region of the two layers
controlled by heterostrain.
In the presence of a lattice modulation between layers, electrons propagating in one layer
can have valley-dependent dissipation due to
valley asymmetric interlayer coupling, thus giving rise to a valley-polarized current.
Such a process can be described by an effective non-Hermitian theory, in which the valley filter is
driven by a valley-resolved non-Hermitian skin effect.
Nearly 100\% valley-polarization can be achieved within a wide parameter range
and the functionality of the valley filter is electrically tunable.
The non-Hermitian topological scenario of the valley filter ensures high
tolerance against imperfections such as disorder and edge defects.
Our work opens a new route for efficient and robust valley filters while
significantly relaxing the stringent implementation requirements.
\end{abstract}

\date{\today}

\maketitle

\emph{Introduction}.-Valleytronics~\cite{Beenakker2007NP,xu2014spin,schaibley2016NRM,vitale2018small},
exploits the valley degree of freedom for electronic information processing and storage
similar to the use of spin degree of freedom in spintronics~\cite{DSarma2004RMP}.
As the \emph{K} and \emph{K'} valleys are distinctly separated in reciprocal space,
graphene and other hexagonal 2D materials have become promising candidates
for the valleytronic devices. Various valley-resolved phenomena
have been explored, including
the valley Hall effect~\cite{NIu2007PRL,yamamoto2015PSJ,gorbachev2014Science,mak2014Science},
optical control of valley polarization~\cite{mak2012NaNo,zeng2012NaNo,cao2012Nc,jones2013NaNo},
and valley Zeeman effect~\cite{srivastava2015NP,aivazian2015NP,LiYilei2014PRL,MacNeill2015PRL,stier2016Nc}.
This demonstrations exemplify the feasibility to manipulate the valley degree of freedom.
Just as spin-polarized current serves as a key ingredient
in spintronics~\cite{DSarma2004RMP}, the implementation of valley-polarized
current is a prerequisite for any valleytronic applications.
Various proposals have been put forward to generate valley-polarized current in graphene,
including using the zeroth mode of the zigzag ribbon~\cite{Beenakker2007NP},
line defects~\cite{GunlyckePRL2011,LiuYangPRB_2013,ChenJHPRB2014}, zero-line modes~\cite{Martin08prl,QiaoZhenhuaPRL2014,qiao2011NanoLetter,li2018valley},
mechanical strain~\cite{fujita2010AIP,JiangPRL2013,zhai2011NJP} and nanobubbles~\cite{Settnes16prl}.
Despite intense efforts, achieving an ideal valley filter capable
of generating fully valley-polarized current with controllable valley
polarization remains challenging~\cite{schaibley2016NRM}.
Therefore, it is crucial to explore new strategies that can achieve
a highly efficient valley filter, and simultaneously, tolerate various device imperfections.

Physical effects that are robust against imperfections are commonly found
in systems with nontrivial topological properties. One such example is the chiral edge states of the
quantum Hall phase, which are robust to disorder~\cite{girvin2019modern}. In Hermitian systems,
nonreciprocal transport can only occur at the boundaries of the sample.
To achieve unidirectional transport in the bulk, it is necessary to
disrupt the Hermiticity of the system~\cite{Lee19prl,Bessho21prl,chen2023fate}, or turn it open.
Recently, great progress has been made in this direction. Unidirectional
transport can be achieved inside the bulk, which causes accumulation of states at
open boundaries, known as the non-Hermitian skin effect~\cite{YaoShunyu2018PRL,YaoShunyu2018PRL2,Kunst2018PRL,
Yokomizo2019PRL,YangZhesen2020PRL,Shao_2022_PRB}. Such nonreciprocal bulk transport
also has a topological origin~\cite{Borgnia2020PRL,Okuma2020PRL,ZhangKai2020PRL,Kunst_RMP_2021},
the point gap topology of the complex spectrum, indicating
its robustness against imperfections. Inspired by the analysis above, a robust
valley filter was suggested to be implemented through valley-resolved nonreciprocal transport in the non-Hermitian regime,
which can be engineered in mesoscopic electronic systems~\cite{Genghao_2023PRB}.

\begin{figure}
\centering
\includegraphics[width=1\columnwidth]{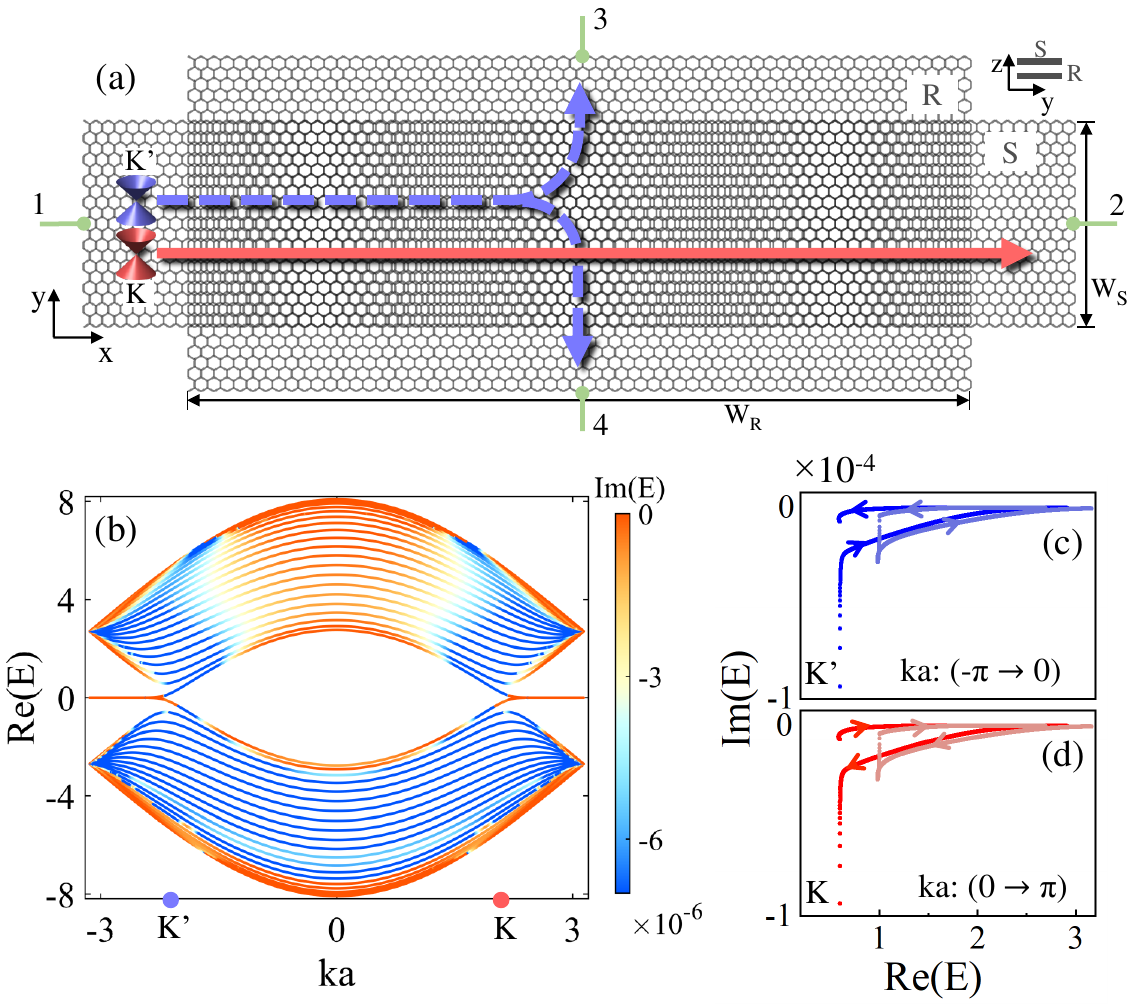}\\
\caption{(a) Schematic diagram of the proposed interface system composed of
an upper graphene layer S and a lower graphene layer R, with four terminals labeled by 1-4.
Electrons in the $K$ valley can transmit from 1 to 2 while
those in the $K'$ valley are filtered out by entering the R layer.
(b) The effective complex bands of S with the interlayer hopping strength
set to one percent of its original value for better illustration.
(c,d) Complex energy spectra of the two lowest sub-conduction bands
(above the bands with zero energy)
with the directions of spectral winding
marked by arrows, which corresponds to the change of $ka$ (c) from $-\pi$ to
$0$ and (d) from $0$ to $\pi$, respectively.
The relevant parameters are $a_0=0.142$ nm, $W_S=29 a_0$, $d_0=0.335$ nm, $t^0_\perp=0.48$ eV and $\delta_0=0.319a_0$.
}\label{fig1}
\end{figure}

In this Letter, we propose a robust valley filter using the 1D moir\'{e} pattern formed
in a graphene homobilayer, as illustrated in Fig.~\ref{fig1}(a), referred to as the moir\'{e} valley filter.
The setup has an interface structure composed of the upper layer (system S) and the lower layer (reservoir R),
along with four terminals (1-4). If layer R is subjected to a uniaxial tension, the coupling between S an R can be
valley-dependent. Due to such a lattice modulation, the electrons in S that flow from terminal 1 to 2
initially occupy both valleys and then undergo a valley selective filtering by R, thus generating
a valley-polarized current [cf. Fig.~\ref{fig1}(a)].
Physically, the valley-dependent coupling with R introduces an effective non-Hermitian self-energy to S,
which leads to valley-resolved nonreciprocal transport.
Specifically, the electrons in the $K$ $(K')$ valley
can only propagate in the $x$ $(-x)$ direction, while its propagation in
the opposite direction will have a strong decay.
The scenario can be interpreted by the valley-resolved complex spectral winding,
which induces nonreciprocity within each valley.
Such non-Hermitian topological effect enables the valley filter to
be resilient against imperfections such as disorder and edge defects,
features of major importance towards its practical implementation.

\emph{Model}.-We take the region S as a zigzag graphene nanoribbon with a width of $W_S$ along
the $x$ direction, stacked on top of the R layer with a length of $W_R$
in such a way that their zigzag edges are aligned with each other. The overlapping area of the two layers is $W_S$ $\times$ $W_R$.
The bilayer system can be captured by the tight-binding Hamiltonian as
\begin{equation}\label{Hamiltonian_original}
\begin{split}
H=-t\sum_{\langle i,j\rangle}c_i^\dag c_j-\sum_{i,j}t_\perp(\mathbf{r}_i,\mathbf{r}_j)c_i^\dag c_j,
\end{split}
\end{equation}
where $c_i^\dag$ $(c_i)$ creates (annihilates) electrons at site $i$.
The first term describes the intralayer nearest-neighbor hopping with strength $t= 2.7$ eV,
and the second term corresponds to the interlayer hopping~\cite{trambly2010NL,UryuPRB2004,nakanishi2001JPJ}
\begin{equation}\label{Hamiltonian_original1}
t_\perp(\mathbf{r}_i,\mathbf{r}_j)=te^{-\frac{r-a_0}{\delta_0}}\left[1-\left(\mathbf{r}\cdot \hat{z}/{r}\right)^2\right]-
t^0_\perp e^{-\frac{r-d_0}{\delta_0}}\left({\mathbf{r}\cdot \hat{z}}/{r}\right)^2.
\end{equation}
Here, $\mathbf{r}=\mathbf{r}_i-\mathbf{r}_j$ is the displacement vector
between two sites,
$\hat{z}$ is the unit vector in the $z$-direction,
$a_0$ is the intralayer site spacing,
$d_0$ is the interlayer spacing,
and $t^0_\perp=0.48$ eV is the vertical hopping between two layers according to the ab initio calculations~\cite{trambly2010NL}.
The hopping decays exponentially within a typical scale $\delta_0$.

The 1D moir\'{e} superlattice~\cite{Timmel20prl}
can be achieved by including a uniaxial strain on the R layer in the $x$-direction.
Accordingly, the scale of R layer is changed by a factor $\varepsilon$ in the $x$-direction
and the $x$-coordinate of the lattice sites changes through $x_i\rightarrow(1+\varepsilon)x_i$.
It is assumed that the $y$-coordinate remains unchanged by neglecting the Poisson's ratio of graphene~\cite{PereiraPRB2009}.
Such deformation of R gives rise to a moir\'{e} pattern in the $x$-direction; see Fig.~\ref{fig1}(a).
The main effect of the deformation of the R layer can be clearly seen in the momentum space,
which drives the two Dirac points $K, K'$ moving towards the $\Gamma$ point
in the projected 1D Brillouin zone~\cite{note}.
As a result, a relative shift takes place between the Dirac cones in the two layers,
which breaks the reciprocity within each valley.
The moir\'{e} superlattice is assumed to be commensurate for simplicity
such that $1+\varepsilon=m/n$ is
a rational number with $m,n\in \mathbb{N}$. The primitive wavevectors of the
S layer, R layer and the whole moir\'{e} superlattice are given by $\mathbf{b}_S=(2\pi/a)\hat{x}$, $\mathbf{b}_R=n\mathbf{b}_S/m$
and $\mathbf{b}_M=\mathbf{b}_S/m$, respectively, with $a=\sqrt{3}a_0$.

\emph{Complex energy bands}.-Although the Hamiltonian~\eqref{Hamiltonian_original}
that describes the bilayer system is inherently Hermitian,
our study instead concentrates
on the non-Hermitian physics within the S layer, whose
electronic property is
engineered by the R layer~\cite{Genghao_2023PRB}.
It provides an important
perspective for the understanding of the transport properties.
Electrons in S can penetrate into R and finally escape into leads 3 and 4,
thus undergoing an effective loss [cf. Fig.~\ref{fig1}(a)].
The physics within S can be described by the Green's function in a general form as
\begin{equation}\label{Hamiltonian_Eff}
\begin{split}
g_S^r(\omega)=\left(\omega-H_S^{\rm{eff}}\right)^{-1},\ \ H_S^{\rm{eff}}(\omega)=H_S+\Sigma^r_R(\omega),
\end{split}
\end{equation}
where the effective Hamiltonian $H_S^{\rm{eff}}$ consists of the bare
Hamiltonian $H_S$ of the S layer and the retarded self-energy $\Sigma^r_R(\omega)$, which is non-Hermitian.
The self-energy is given by $\Sigma^r_R(\omega)=\hat{V}g_R^r(\omega)\hat{V}^{\dagger}$, where $\hat{V}$ is the interlayer
hopping matrix and $g_R^r(\omega)=\left(\omega-H_R-\Sigma^r_3-\Sigma^r_4\right)^{-1}$ is the retarded Green's
function of the R layer. Here, $H_R$ is the Hamiltonian of the R layer and the effect of leads 3 and 4 is absorbed into the surface self-energy
$\Sigma^r_{3,4}$, which is calculated using the iterative method~\cite{sm}.

The electron dynamics in S is governed by $H_S^{\rm{eff}}(\omega)$, making it insightful
to explore its complex energy bands. For this purpose, we first assume that
both S and R layers extend to infinity in the $x$-direction.
For the commensurate moir\'{e} modulation, the interlayer coupling takes place
between the Bloch states with their wave vectors differing by a reciprocal superlattice vector
$\mathbf{G}_\alpha =l\mathbf{b}_M$, where $\alpha=S,R$ is the layer index and $l\in \mathbb{Z}$.
The Bloch Hamiltonian for the bilayer is expressed as
\begin{equation}\label{HK}
\begin{split}
H(\mathbf{k})&=\sum_{\alpha,\beta=S,R}\sum_{\mathbf{G}_\alpha,\mathbf{G}'_\beta}c^{ \dag}_{\mathbf{k}+\mathbf{G}_\alpha} H_{\mathbf{G}_\alpha,\mathbf{G}'_\beta}(\mathbf{k})c_{\mathbf{k}+\mathbf{G}'_\beta},
\end{split}
\end{equation}
where the Bloch wave vector $\mathbf{k}\in[-\mathbf{b}_M/2,\mathbf{b}_M/2]$ lie in the moir\'{e} Brillouin zone
and the summation over the reciprocal superlattice vector is restricted by
the condition that the total wave vector $\textbf{k}+\textbf{G}_\alpha=\textbf{q}_\alpha$
lies in the first Brillouin zone of the $\alpha$ layer. As such, the reciprocal vectors $\{\textbf{G}_\alpha\}$ become
band labels and all bands are folded up into the moir\'{e} Brillouin zone. The Fermi annihilation operator is expressed by the sublattice components as $c_{\mathbf{q}_\alpha}=(c^{1}_{\mathbf{q}_\alpha},c^{2}_{\mathbf{q}_\alpha},\cdots,c^{\chi}_{\mathbf{q}_\alpha},\cdots)^{\text{T}}$
with $c^{\chi}_{\mathbf{q}_\alpha}=
N_\alpha^{-1/2}\sum_{\mathbf{r}_\alpha^{\chi}}
e^{i{\mathbf{q}_\alpha}\cdot\mathbf{r}_\alpha^{\chi}}
c_{\mathbf{r}_{\alpha}^{\chi}}$,
where the lattice site is specified by its location $\mathbf{r}_\alpha^{\chi}$
and $N_\alpha$ is the number of unit cells.
The matrix elements in Eq.~\eqref{HK} are specified as follows.
$H_{\textbf{G}_S,\textbf{G}'_S}=H_S(\textbf{k}+\textbf{G}_S)\delta_{\textbf{G}_S,\textbf{G}'_S}$ is diagonal in the $\textbf{G}_S$ space
with $H_S(\textbf{k}+\textbf{G}_S)$ the Bloch Hamiltonian of the S layer.
Similarly, $H_{\textbf{G}_R,\textbf{G}'_R}=\tilde{H}_R(\textbf{k}+\textbf{G}_R)\delta_{\textbf{G}_R,\textbf{G}'_R}$
is diagonal in the $\textbf{G}_R$ space with $\tilde{H}_R(\mathbf{k}+\mathbf{G}_R)
= H_R(\mathbf{k}+\mathbf{G}_R) + \Sigma^r_3(\mathbf{k}+\mathbf{G}_R) + \Sigma^r_4(\mathbf{k}+\mathbf{G}_R)$ the effective Hamiltonian of the R layer
composed of the bare Hamiltonian and the self-energies due to leads 3 and 4~\cite{sm}.
Finally, the interlayer coupling matrix is calculated by $H_{\textbf{G}_S,\textbf{G}'_R}=H^\dag_{\textbf{G}'_R,\textbf{G}_S}
=\langle\Phi_{\mathbf{k}+\mathbf{G}_S}^{S}|\hat{V}|\Phi_{\mathbf{k}+\mathbf{G}'_R}^{R}\rangle$,
in which $|\Phi_{\mathbf{q}_\alpha}^{\alpha}\rangle$ is the Bloch function of the $\alpha$ layer.

Integrating out the R layer following Eq.~\eqref{Hamiltonian_Eff} yields
the self-energy matrix $\Sigma^r_R(\omega,\textbf{k};\textbf{G}_S,\textbf{G}'_S)
=\sum_{\textbf{G}_R}H_{\textbf{G}_S,\textbf{G}_R}[\omega-\tilde{H}_R(\textbf{k}+\textbf{G}_R)]^{-1}H^\dag_{\textbf{G}'_S,\textbf{G}_R}$.
The complex band structures can be obtained by diagonalizing the effective Hamiltonian of the S layer
$H_S^{\text{eff}}(\omega,\textbf{k})=H_S+\Sigma^r_R$. Due to the folding of the complex bands,
the information regarding the two valleys is hidden. To extract the valley flavor of the states, we unfold the
multiple bands back to the original Brillouin zone of the S layer by tracking the momentum
index $\mathbf{G}_S$~\cite{note2}. Futhermore, since the self-energy $\Sigma_R^r(\omega, \mathbf{k})$
is itself a function of energy $\omega$,
only the eigenvalues of $H_S^{\rm{eff}}(\omega,\textbf{k})$ that fulfill $\text{Re}[E(\mathbf{k})]=\omega$
have physical meanings and are considered.
The complex band structure, shown in Fig.~\ref{fig1}(b),
highlights both the real and imaginary parts of the energy spectrum,
for which the valley-resolved nonreciprocal transport can be clearly revealed.
Specifically, the lifetime $\tau(\omega, \mathbf{k})$ of electrons in S
is determined by $\tau^{-1}=-\text{Im}[E(\omega, \mathbf{k})]$. From Fig.~\ref{fig1}(b).
It is observed that the right-moving electrons around the $K$ valley possess a longer lifetime
compared to the left-moving electrons within the same valley.
The opposite situation applies to
the $K'$ valley as expected from time-reversal symmetry, thus exhibiting
a valley-resolved nonreciprocity.
Such scenario can be further illustrated
by the complex spectral winding~\cite{Borgnia2020PRL,Okuma2020PRL,ZhangKai2020PRL,Kunst_RMP_2021}.
Here, we have separated the spectral winding for the two valleys depicted in Figs.~\ref{fig1}(c) and~\ref{fig1}(d),
where the arrows represent the momentum increase.
Note that only the two lowest sub-conduction bands are shown for clarity. The complex energy spectra
in Figs.~\ref{fig1}(c) and~\ref{fig1}(d) are identical, but they exhibit opposite
winding directions, indicating a valley-resolved non-Hermitian skin effect.
The spectral winding, along with the distinct separation of the two valleys
in the momentum space, indicates the robustness of this effect.

\begin{figure}[!htb]
\centering
\includegraphics[width=1\columnwidth]{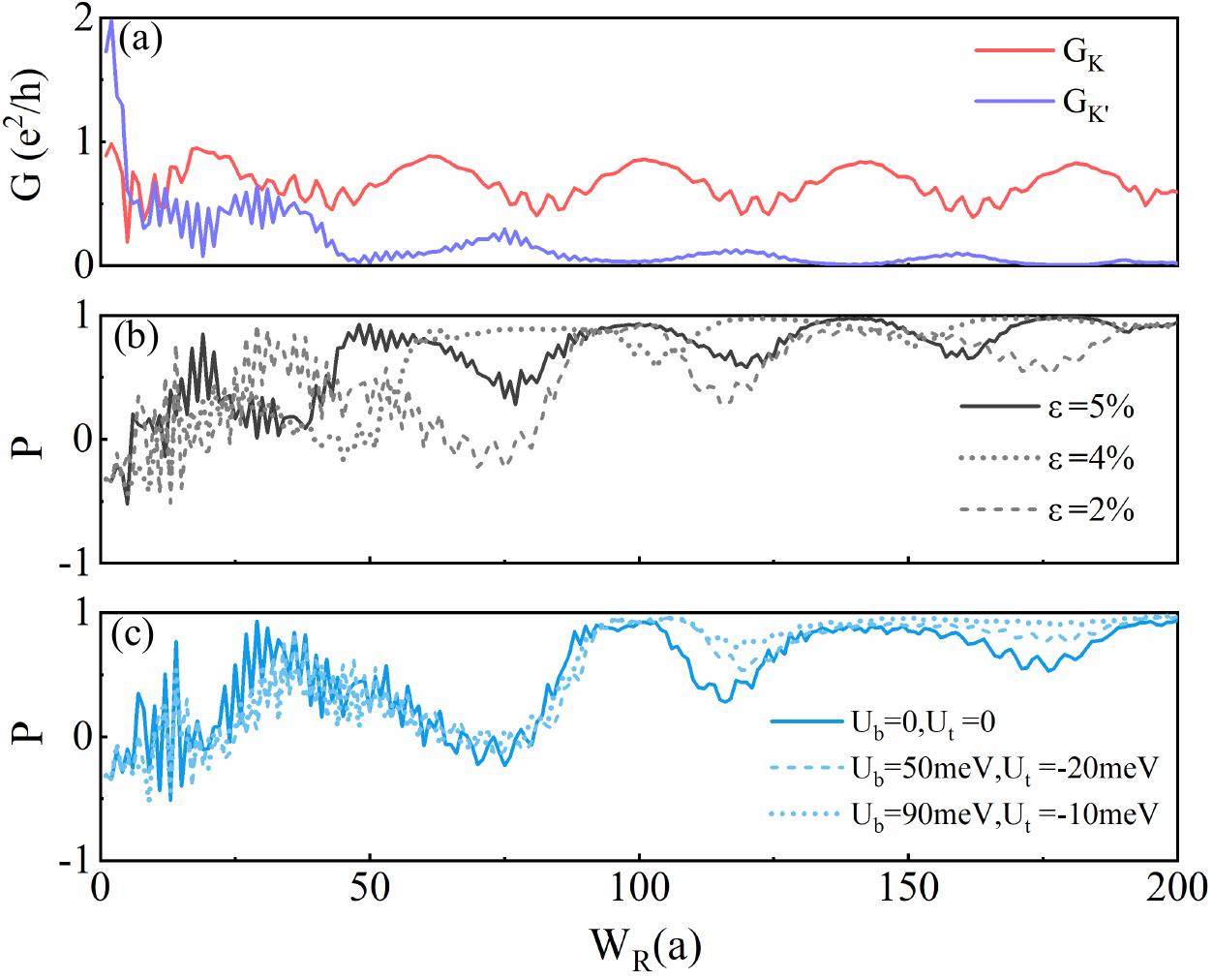}\\
\caption{(a) Valley-resolved differential conductance as a function of $W_R$ with $\varepsilon=5\%$.
(b) Valley polarization with different distortion factors $\varepsilon$.
(c) Valley polarization adjusted by the gate voltages with $\varepsilon=2\%$.
The bias voltage is $eV=0.81$ eV and the other parameters are the same as those in Fig.~\ref{fig1}.
}\label{fig2}
\end{figure}

\emph{Non-Hermitian moir\'{e} valley filter}.-The valley-resolved nonreciprocal transport
can be utilized to create a novel type of valley filter. In contrast to previous
proposals, the current scenario exploits a non-Hermitian strategy.
We numerically study the electron transport between lead 1 and 2 through the
moir\'{e} valley filter using the standard scattering
matrix approach~\cite{datta1997electronic} based on the Hermitian Hamiltonian~\eqref{Hamiltonian_original}.
To reveal the valley-resolved scattering, we label different transport channels with the
wave vector $\textbf{k}$ for a specific incident energy $\omega$~\cite{sm}.
In particular, the amplitude $t_{\mathbf{k}_2,\mathbf{k}_1}(\omega)$ corresponds to the transmission from
channel $\mathbf{k}_1$ in lead 1 into channel $\mathbf{k}_2$ in lead 2.
Within the energy window of interest where the valley degree of freedom is well defined,
electrons that lie in the $K$ and $K'$ valley can be
selected by the conditions of $k\in[0,\pi/a]$ and $k\in[-\pi/a,0]$, respectively~\cite{sm}.

\begin{figure}
\centering
\includegraphics[width=\columnwidth]{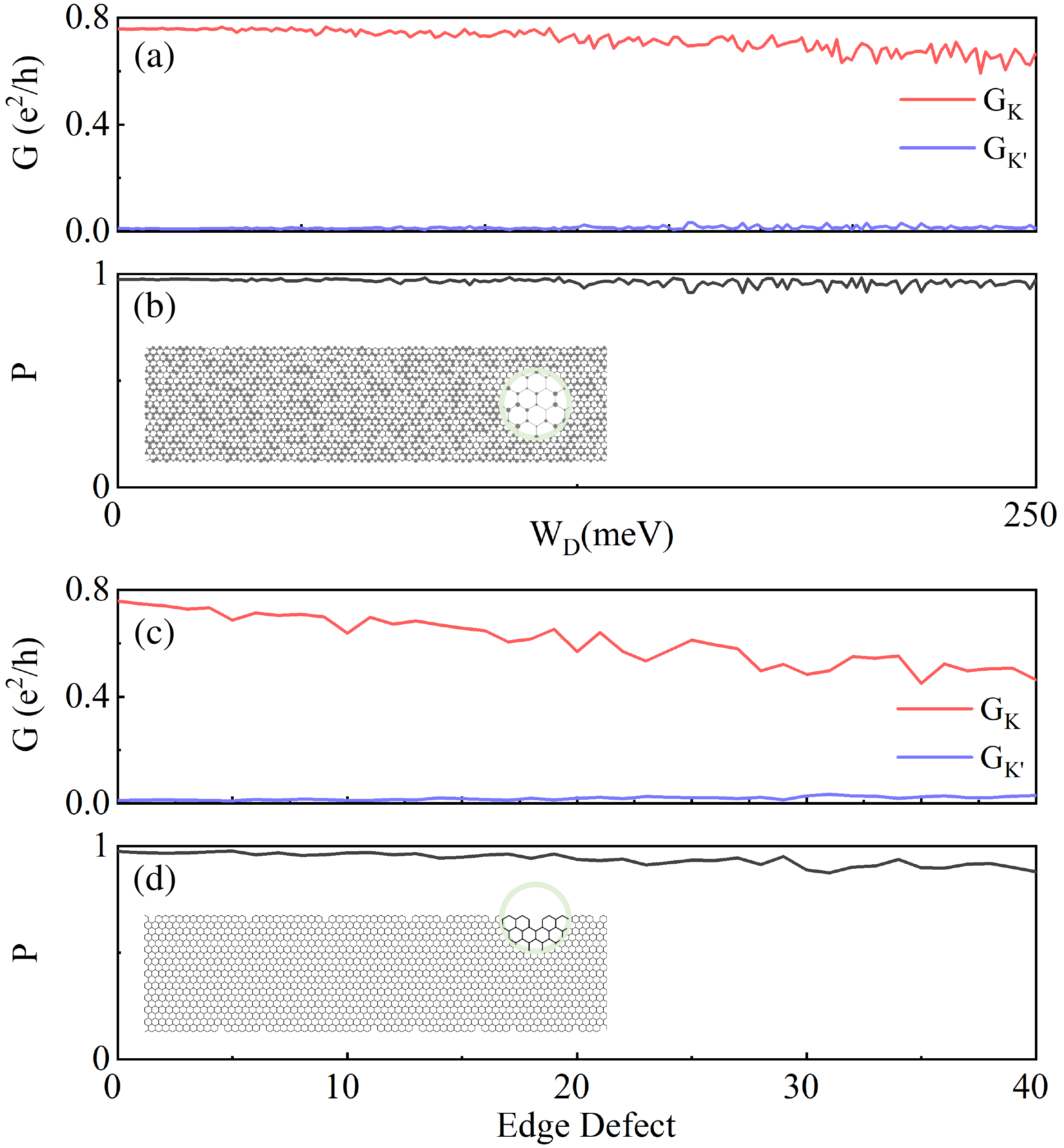}\\
\caption{Valley-resolved conductance and valley polarization as a function of
(a,b) disorder strength and (c,d) number of edge defects.
The relevant parameters are $\varepsilon=5\%$, $W_R=141 a$, and the other parameters are the same as those in Fig.~\ref{fig1}.
}\label{fig3}
\end{figure}

By applying a bias voltage $V$ on lead 1 and keeping the other three leads grounded,
the differential conductance with the valley index $\kappa=K,K'$ is defined as
\begin{equation}\label{TkTkp}
\begin{split}
G_\kappa(eV)=\frac{e^2}{h}\sum_{\mathbf{k}_1}\sum_{\mathbf{k}_2\in \kappa}|t_{\mathbf{k}_2,\mathbf{k}_1}|^2,
\end{split}
\end{equation}
which is contributed by the transmitted electrons in the $\kappa$ valley
in lead 2, while the incident electrons from lead 1 occupy both valleys. The numerical results
of $G_{K}$ and $G_{K'}$ as a function of $W_R$ are shown in Fig.~\ref{fig2}(a).
Because of the valley-resolved nonreciprocity, $G_{K'}$ decays much faster than $G_{K}$ as $W_R$ increases.
Furthermore, both conductances exhibit additional oscillations arising from the intravalley interference.

The valley polarization of the current is quantified by~\cite{Beenakker2007NP}
\begin{equation}
P=\frac{G_{K}-G_{K'}}{G_{K}+G_{K'}},
\end{equation}
where $P>0$ ($P<0$) correspond to valley polarization at $K$ ($K'$).
The valley polarization corresponding to the conductance in Fig.~\ref{fig2}(a) is depicted in Fig.~\ref{fig2}(b) by the
solid line. It can be observed that the current becomes $K$-valley polarized as $W_R$
exceeds a certain value. Notably, a polarization of $P\simeq 100\%$ can be achieved around the plateaus.
The results with different deformation factors $\varepsilon$ are compared in Fig.~\ref{fig2}(b),
where the contrast in $P$ that appears first for small $W_R$ decreases and eventually disappears as $W_R$ increases.
A similar situation occurs for the setup with different widths $W_S$~\cite{sm},
indicating that the decrease of the valley polarization for a larger $W_S$ can be
compensated by a larger $W_R$. These results
highlight the universality of our proposal. The positive valley
polarization here underscores the main distinction between the current scenario
and the one in Ref.~\cite{Beenakker2007NP}. For the latter, the presence of an additional
right-moving mode at the $K'$ valley in the conduction band would lead
to a negative polarization with the same parameters.

The valley polarization can be further enhanced by applying gate voltages~\cite{oostinga2008gate,Castro07prl,chen2020NC,sui2015NP,shimazaki2015NP}.
Their effects can be simulated by two on-site potential $U_{t}$ and $U_{b}$ in the respective
layer as $H_g=\sum_{i\in S}U_{t}c_i^\dag c_i+\sum_{i\in R}U_{b}c_i^\dag c_i$. The impact of the gate voltages
is illustrated in Fig.~\ref{fig2}(c), which results in a suppression of the oscillation in $P$
and effectively expands the parameter region with $P\simeq 100\%$.
This property alleviates the stringent requirement for a specific $W_R$ to achieve an efficient valley filter.

\emph{Robustness.-}In contrast to previous proposals~\cite{Beenakker2007NP,GunlyckePRL2011,LiuYangPRB_2013,ChenJHPRB2014,Martin08prl,QiaoZhenhuaPRL2014,qiao2011NanoLetter,li2018valley,fujita2010AIP,JiangPRL2013,zhai2011NJP,Settnes16prl},
our scheme capitalizes on the non-Hermitian topological effect, making it more robust
against various lattice imperfections.
Consider an electron in the $K$ valley incident from lead 1. It may be scattered to the $K'$ valley, either
with or without changing its propagating
direction. In the case of inter-valley forward scattering, the electron rapidly decays
into the R layer and exhibits a short lifetime, as shown in Fig.~\ref{fig1}(b),
preventing it from reaching lead 2. As for the inter-valley backward scattering, the electron is
reflected back into lead 1. Therefore, both two processes of
inter-valley scattering do not contribute to the output current in lead 2,
and so leave the valley polarization unaffected.

Here, we examine two types of imperfections commonly encountered in
the system: local disorder and edge defects.
We first consider the effect of Anderson disorder depicted by the inset of Fig.~\ref{fig3}(b).
It is described by the random on-site potential $H_{\text{dis}}=\sum_{i\in S}U_i c_i^\dag c_i$
in the S layer, where $U_i\in[-W_D/2, W_D/2]$ with $W_D$ the disorder strength.
The valley-resolved transport is shown in Figs.~\ref{fig3}(a) and \ref{fig3}(b), in which
the high valley polarization persists despite the disorder of a moderate strength.
Moving on to the edge defects, we consider vacancies that randomly exist
at the zigzag edges of the S layer; refer to the inset of
Fig.~\ref{fig3}(d).
Similar to the disorder effect, the valley polarization remains high even
in the presence of a large number of edge defects as shown in Figs.~\ref{fig3}(c) and \ref{fig3}(d).
Moreover, the impact of edge defects is expected to diminish as the width $W_S$ increases
because the valley filter relies primarily on bulk transport.

During the fabrication a small twist between the two layers can appear accidentally.
This is simulated by a rotation of the R layer and
consequently, the interlayer hopping changes its magnitude according to Eq.~\eqref{Hamiltonian_original1}
by using the twisted geometry. By calculating the conductance $G_{K,K'}$ and valley polarization $P$ as
a function of the twist angle, we find that the valley polarization is always positive and exhibits an
interesting oscillation behavior. Furthermore, nearly perfect
valley polarization can be realized around different plateaus; see Supplemental Material~\cite{sm}
for more details.

\emph{Discussion and outlook.-}We have demonstrated an efficient strategy to create a valley filter in a
heterostrained graphene bilayer stemming from the moir\'{e} modulation.  The
valley filter effect relies on the 1D moir\'{e} pattern, which can
be realized by a differential and uniaxial
strain~\cite{si2016strain,bai2020NM}.  The underlying non-Hermitian topological
scenario guarantees its robustness against lattice imperfections, lifting the
stringent implementation requirements of previous proposals.  The universality
of the mechanism implies that the reduction of nonreciprocity due to increased
$W_S$ can be compensated by using a larger $W_R$. Moreover,
the device with a larger scale effectively reduces the energy spacing between
transverse modes and so a small bias voltage is sufficient to
drive the valley-resolved transport. Compared
with previous proposals using valley polarized edge transport~\cite{pan14prl,pan15prb,zhou16prb},
our scheme can generate a larger valley current and can be directly applied to
construct valleytronic devices, such as the valley valve.
Due to the bulk origin of
our mechanism, the impact of irregular edges becomes negligible for wide
devices.  Our moir\'{e} valley filter mechanism can be readily extended to
other hexagonal 2D materials, including transition metal
dichalcogenides~\cite{schaibley2016NRM,vitale2018small},
where the spin-valley locking effect can
further enhance the robustness of the valley filter
against nonmagnetic disorders~\cite{zhou21prl}.  Furthermore, a
similar strategy can be applied to create the valley-resolved non-Hermitian
skin effect in metamaterials, such as photonic and phononic
crystals~\cite{weidemann2020topological,ghatak2020observation,zhang2021observation,zhang2021acoustic,zhou2023observation}.
Our results pave the way for exploring the valley-resolved non-Hermitian skin effect,
establishing an overlooked connection between non-Hermiticity and
moir\'{e} physics~\cite{andrei2020NM,balents2020superconductivity,andrei2021marvels,kennes2021moire,lau2022Nature,mak2022semiconductor}.

\begin{acknowledgments}
This work was supported by the
National Key Projects for Research and Development of China under Grant No. 2022YFA120470 (W.C.),
the National Natural Science Foundation of
China under Grant No. 12074172 (W.C.), No.  12222406  (W.C.) and
No. 12174182 (D.Y.X.), and the State Key Program for Basic
Researches of China under Grants No. 2021YFA1400403 (D.Y.X.).
J.L.L acknowledges the support from Aalto Science-IT project,
the Academy of Finland Projects No. 331342,
No. 336243 and No. 358088, and the Jane and Aatos Erkko Foundation.
\end{acknowledgments}


\begin{thebibliography}{83}
\bibitem [{not({\natexlab{a}})}]{note}%
    \BibitemOpen
    \href@noop {} {} \bibinfo {note} {The deformation of the R
    layer slightly changes the intralayer nearest-neighbor hopping as well. This
    modification of intralayer hopping is also governed by
    Eq.~\eqref{Hamiltonian_original1}, but its impact is negligible.}\BibitemShut
    {Stop}%
\bibitem [{not({\natexlab{b}})}]{note2}%
    \BibitemOpen
    \href@noop {} {}\ \bibinfo {note} {During the band-folding
    process, one can establish a correspondence between the band index $n$ in the
    moir\'{e} Brillouin zone and the momentum shift
    $\mathbf{G}_S^n=n\mathbf{b}_M$ of the state in the original Brillouin zone of
    S, where $n$ is an integer. Band unfolding can be understood as the inverse
    process~\cite{Boykin05prb,Popescu10prl,Medeiros14prb}. When coupling between
    different bands occurs in the moir\'{e} Brillouin zone, we solve the
    eigenstates and identify the component $n'$ with the highest weight and
    unfold the band through $E_{n'}(\mathbf{k})=E(\mathbf{k}+\mathbf{G}_S^{n'})$.
    The complete band unfolding is achieved by repeating this procedure for all
    bands in the moir\'{e} Brillouin zone. By doing so, the valley-associated
    information, initially smeared in the moir\'{e} Brillouin zone, is
    recovered.}\BibitemShut {Stop}%
\end{thebibliography}

%

\end{document}